# Stability of Vertical Steady Rotation of an Ellipsoid On a Smooth Horizontal Plane

Milan Batista

(Dec 2006)

**Abstract**

The article treats the classical problem of stability of steady rotation of a rigid homogeneous ellipsoid on a rigid smooth plane which rotates about its vertical axis. The condition for the steady rotation is derived from the Euler-Poisson equations.

## 1. Introduction

The problem which will be discussed follows (Routh [1], pp 202 §252 Example. 2): '*An ellipsoid is placed with one of its vertices in contact with a smooth horizontal plane. What angular velocity of rotation must it have about the vertical axis in order that the equilibrium may be stable?*'

As was noted by Routh ([1],) the problem was solved by M.Puiseux in 1852 and he gives result: '*Let a, b, c be the semi-axes, c the vertical axis, then the angular velocity must be grater than* $\sqrt{\frac{5g}{c}\frac{\sqrt{c^4-a^4}+\sqrt{c^4-b^4}}{a^2+b^2}}$'. The details of how the solution was obtained are not provided except the note that M.Puiseux derives it from Euler-Poisson equations.

While the interested reader can probably obtain the solution from the original source, it is surprising that (to the author's knowledge) the solution of the problem can not be found in any standard textbook (historical or contemporary) dealing with rigid body dynamics neither as example nor as an exercise (with the exception of Routh's book). Yet the problem is a good illustration of a holonomic system that has a relatively simple analytical solution. The purpose of the present article is to provide the detailed solution of the problem.

## 2. Equations

Consider that the homogeneous ellipsoid of mass *m* rolls on a smooth horizontal plane under the influence of the uniform gravitational field characterized by the acceleration *g*. Let $\{\hat{\boldsymbol{e}}_1, \hat{\boldsymbol{e}}_2, \hat{\boldsymbol{e}}_3\}$ be the inertial reference base fixed on a plane with $\hat{\boldsymbol{e}}_3$ vertical and coordinates $(x_1, x_2, x_3)$ and $\{\hat{\boldsymbol{\varepsilon}}_1, \hat{\boldsymbol{\varepsilon}}_2, \hat{\boldsymbol{\varepsilon}}_3\}$ the body-fixed base with coordinates $(\xi_1, \xi_2, \xi_3)$ which have the directions of the ellipsoid principle axes of inertia and origin in the centre of the mass point of the ellipsoid. The connection between inertial and body coordinates is given by

$$x_i = \bar{x}_i + \sum_{j=1}^{3} \alpha_{ji} \xi_j \tag{1}$$

where $\bar{x}_i$ $(i = 1,2,3)$ are coordinates of the centre of the mass point and $\alpha_{ij} = \hat{\boldsymbol{\varepsilon}}_i \cdot \hat{\boldsymbol{e}}_j$ $(i, j = 1,2,3)$ are directional cosines.

*The contact point.* Let the surface of the ellipsoid be given by the equation

$$f = \left(\frac{\xi_1}{a_1}\right)^2 + \left(\frac{\xi_2}{a_2}\right)^2 + \left(\frac{\xi_3}{a_3}\right)^2 - 1 = 0 \tag{2}$$

where $a_1$, $a_2$ and $a_3$ are ellipsoid semi-axes. From (1) the equation of the plane is

$$x_3 = \bar{x}_3 + \alpha_{13}\xi_1 + \alpha_{23}\xi_2 + \alpha_{33}\xi_3 = 0 \tag{3}$$

To calculate the coordinates of the contact point and vertical position of the centre of mass point two more equations are needed. These are obtained from the requirement that at contact point the normal $\boldsymbol{n}$ to the ellipsoid surface is parallel to the normal plane;

i.e., $\boldsymbol{n} = \lambda \hat{\boldsymbol{e}}_3$ where $\lambda$ is an unknown parameter[1]. From (2) the normal $\boldsymbol{n}$ can be expressed as $\boldsymbol{n} = \frac{\xi_1}{a_1^2}\hat{\varepsilon}_1 + \frac{\xi_2}{a_2^2}\hat{\varepsilon}_2 + \frac{\xi_3}{a_3^2}\hat{\varepsilon}_3$ so $\boldsymbol{n}\cdot\hat{\boldsymbol{e}}_3 = \frac{\xi_1}{a_1^2}\alpha_{13} + \frac{\xi_2}{a_2^2}\alpha_{23} + \frac{\xi_3}{a_3^2}\alpha_{33}$. From this follows the equation

$$\frac{\xi_1}{a_1^2}\alpha_{13} + \frac{\xi_2}{a_2^2}\alpha_{23} + \frac{\xi_3}{a_3^2}\alpha_{33} = \lambda \tag{4}$$

Since $\alpha_{13}, \alpha_{23}, \alpha_{33}$ are components of the unit vector one also has

$$\alpha_{13}^2 + \alpha_{23}^2 + \alpha_{33}^2 = 1 \tag{5}$$

Comparing (4) and (5) one finds that the solution of (4) may be sought in the form[2]

$$\xi_1 = \lambda a_1^2 \alpha_{13} \qquad \xi_2 = \lambda a_2^2 \alpha_{23} \qquad \xi_3 = \lambda a_3^2 \alpha_{33} \tag{6}$$

By substituting (6) into (3) one finds the position of the centre of the mass

$$\overline{x}_3 = -\lambda\left(a_1^2\alpha_{13}^2 + a_2^2\alpha_{23}^2 + a_1^2\alpha_{33}^2\right) \tag{7}$$

To determine $\lambda$ the solution (6) is substituted into (2). In this way one finds $\lambda^2\left(a_1\alpha_{13}^2 + a_2\alpha_{23}^2 + a_3\alpha_{33}^2\right) = 1$ and from this, by observing that in (7) one should have $\overline{x}_3 > 0$, one gets

$$\lambda = -\frac{1}{\sqrt{a_1\alpha_{13}^2 + a_2\alpha_{23}^2 + a_3\alpha_{33}^2}} \tag{8}$$

The vertical position of the centre of the mass in the inertial frame is thus from (7) given by

[1] One can of course normalize the normal in advance.

[2] To avoid notation explosion the coordinates of the contact point will be denoted as $(\xi_1, \xi_2, \xi_3)$.

$$\boxed{\overline{x}_3 = \sqrt{a_1^2\alpha_{13}^2 + a_2^2\alpha_{23}^2 + a_3^2\alpha_{33}^2}} \tag{9}$$

and the coordinates of the contact point with the plane in body fixed base are from (6)

$$\boxed{\begin{gathered} \xi_1 = -\frac{a_1^2\alpha_{13}}{\sqrt{a_1^2\alpha_{13}^2 + a_2^2\alpha_{23}^2 + a_3^2\alpha_{33}^2}} \qquad \xi_2 = -\frac{a_2^2\alpha_{23}}{\sqrt{a_1^2\alpha_{13}^2 + a_2^2\alpha_{23}^2 + a_3^2\alpha_{33}^2}} \\ \xi_3 = -\frac{a_3^2\alpha_{33}}{\sqrt{a_1^2\alpha_{13}^2 + a_2^2\alpha_{23}^2 + a_3^2\alpha_{33}^2}} \end{gathered}} \tag{10}$$

These formulas are obtained by Poisson in a slightly different way (Poisson [3] pp 184)

*Dynamics.* The forces acting on the ellipsoid are the gravity and the reaction force at contact point. The later is given by

$$\boldsymbol{F} = F\,\hat{\boldsymbol{e}}_3 = F\left(\alpha_{13}\hat{\boldsymbol{\varepsilon}}_1 + \alpha_{23}\hat{\boldsymbol{\varepsilon}}_2 + \alpha_{33}\hat{\boldsymbol{\varepsilon}}_3\right) \tag{11}$$

The magnitude of the reaction force $F$ is calculated by Newton's law

$$\boxed{F = mg + m\frac{d^2\overline{x}_3}{dt^2}} \tag{12}$$

Because all the forces acts in the vertical direction, the remaining equations for movement of the centre of the mass are

$$m\frac{d^2\overline{x}_1}{dt^2} = 0 \qquad m\frac{d^2\overline{x}_2}{dt^2} = 0 \tag{13}$$

These equations show that the projection of the centre of the mass point on the plane moves in a straight line or stands still.

The reaction force produces with respect to the body centre of the mass the moment which is in the body-fixed frame given by

$$\boldsymbol{M} = F\left(\alpha_{33}\xi_2 - \alpha_{23}\xi_3\right)\hat{\boldsymbol{\varepsilon}}_1 + F\left(\alpha_{13}\xi_3 - \alpha_{33}\xi_1\right)\hat{\boldsymbol{\varepsilon}}_2 + F\left(\alpha_{23}\xi_1 - \alpha_{13}\xi_2\right)\hat{\boldsymbol{\varepsilon}}_3 \tag{14}$$

Substituting (14) into Euler's equations of motion (Goldstein [1], pp205 Eq 5-39) one obtains the Euler-Poisson equations (Poisson [3], pp 186)

$$\boxed{\begin{aligned} I_1\frac{d\omega_1}{dt} + \left(I_3 - I_2\right)\omega_2\omega_3 &= F\left(\alpha_{33}\xi_2 - \alpha_{23}\xi_3\right) \\ I_2\frac{d\omega_2}{dt} + \left(I_1 - I_3\right)\omega_3\omega_1 &= F\left(\alpha_{13}\xi_3 - \alpha_{33}\xi_1\right) \\ I_3\frac{d\omega_3}{dt} + \left(I_2 - I_1\right)\omega_1\omega_2 &= F\left(\alpha_{23}\xi_1 - \alpha_{13}\xi_2\right) \end{aligned}} \tag{15}$$

where the principal moments of inertia of the ellipsoid are (Synge [4], pp 292)

$$I_1 = \frac{m}{5}\left(a_2^2 + a_3^2\right) \qquad I_2 = \frac{m}{5}\left(a_3^2 + a_1^2\right) \qquad I_3 = \frac{m}{5}\left(a_1^2 + a_2^2\right) \tag{16}$$

*Stedy vertical rotation.* For steady vertical rotation with $\xi_3$ axis upward one should have

$$\omega_1 = \omega_2 = 0 \qquad \omega_3 = \omega_0 \tag{17}$$

By this, (10) and the requirement that $F \neq 0$ the system (15) reduces to

$$0 = \alpha_{23}\alpha_{33} \qquad 0 = \alpha_{13}\alpha_{33} \qquad 0 = \alpha_{13}\alpha_{23} \tag{18}$$

From this and (5) one finds

$$\alpha_{13} = \alpha_{23} = 0 \qquad \alpha_{33} = 1 \tag{19}$$

The rest of the unknowns are now obtained from (9), (10) and (12)

$$\bar{x}_3 = a_3 \qquad \xi_1 = \xi_2 = 0 \qquad \xi_3 = -a_3 \qquad F = mg$$

**3. Stability of steady motion**

The classical analysis of the stability of steady motion is based on consideration of steady motion perturbation. To perform this, first the angular orientation of the ellipsoid must be specified. Here for the angular orientation the Bryant angles will be used, which are obtained by the consecutive rotations through the angle $\theta_1$ about the axis $\xi_1$, then through the angle $\theta_2$ about intermediary $\xi_2$ and finally through the angle $\theta_3$ about intermediary $\xi_3$ axis (see Appendix). In the present case, however, because the ellipsoid uniform rotates about the vertical, the rotation about $\xi_3$ axis is through the angle $\omega_0 t + \theta_3$.

Now, if $\theta_1$, $\theta_2$, and $\theta_3$ are small angles--i.e. if $\theta_1, \theta_2, \theta_3 = O(\varepsilon)$ where--then it follows from (A34) that

$$\alpha_{13} = \theta_1 \sin\omega_0 t - \theta_2 \cos\omega_0 t + O\left(\varepsilon^2\right) \qquad \alpha_{23} = -\theta_1 \sin\omega_0 t + \theta_2 \cos\omega_0 t + O\left(\varepsilon^2\right)$$
$$\alpha_{33} = 1 + O\left(\varepsilon^2\right) \tag{20}$$

By this one has from (9) and (12)

$$\bar{x}_3 = a_3 + O\left(\varepsilon^2\right) \qquad F = mg + O\left(\varepsilon^2\right) \tag{21}$$

Further, the components of angular velocity in the body fixed base are, by using (A35), approximated as follows

$$\omega_1 = \frac{d\theta_1}{dt}\cos\omega_0 t + \frac{d\theta_2}{dt}\sin\omega_0 t + O\left(\varepsilon^2\right)$$

$$\omega_2 = -\frac{d\theta_1}{dt}\sin\omega_0 t + \frac{d\theta_2}{dt}\cos\omega_0 t + O\left(\varepsilon^2\right) \tag{22}$$

$$\omega_3 = \omega_0 + \frac{d\theta_3}{dt} + O\left(\varepsilon^2\right)$$

By this system (15) becomes

$$\frac{d^2\theta_1}{dt^2}\cos\omega_0 t + \frac{d^2\theta_2}{dt^2}\sin\omega_0 t - \frac{2a_2^2\omega_0^2}{a_2^2+a_3^2}\left(\frac{d\theta_1}{dt}\sin\omega_0 t - \frac{d\theta_2}{dt}\cos\omega_0 t\right) - \frac{5g\left(a_3^2-a_2^2\right)}{a_3\left(a_2^2+a_3^2\right)}\left(\theta_1\cos\omega_0 t + \theta_2\sin\omega_0 t\right) = 0$$

$$-\frac{d^2\theta_1}{dt^2}\sin\omega_0 t + \frac{d^2\theta_2}{dt^2}\cos\omega_0 t - \frac{2a_1^2\omega_0^2}{a_1^2+a_3^2}\left(\frac{d\theta_1}{dt}\cos\omega_0 t + \frac{d\theta_2}{dt}\sin\omega_0 t\right) - \frac{5g\left(a_3^2-a_1^2\right)}{a_3\left(a_1^2+a_3^2\right)}\left(\theta_1\sin\omega_0 t - \theta_2\cos\omega_0 t\right) = 0 \tag{23}$$

$$\frac{d^2\theta_3}{dt^2} = 0$$

From the last equation $\theta_3 = \theta_{30} + \gamma t$ and the inspection of the remaining two equations suggest defining the new functions

$$\phi_1 \equiv \theta_1\cos\omega_0 t + \theta_2\sin\omega_0 t \qquad \phi_2 \equiv \theta_1\sin\omega_0 t - \theta_2\cos\omega_0 t \tag{24}$$

By this the first two equations of (23) transform to the system of linear equations with constant coefficients for unknowns $\phi_1$ and $\phi_2$ of the form

$$\frac{d^2\phi_1}{dt^2}-\frac{\left(a_3^2-a_2^2\right)}{\left(a_2^2+a_3^2\right)}\left(\omega_0^2+\frac{5g}{a_3}\right)\phi_1+\frac{2a_2^2\omega_0^2}{a_2^2+a_3^2}\frac{d\phi_2}{dt}=0$$

$$\frac{d^2\phi_2}{dt^2}-\frac{\left(a_3^2-a_1^2\right)}{\left(a_1^2+a_3^2\right)}\left(\omega_0^2+\frac{5g}{a_3}\right)\phi_2-\frac{2a_1^2\omega_0^2}{a_1^2+a_3^2}\frac{d\phi_1}{dt}=0 \qquad (25)$$

*The characteristic equation.* The solution of (25) is sought in the form

$$\phi_1=C_1e^{\lambda t}\qquad \phi_2=C_2e^{\lambda t} \qquad (26)$$

where $C_1$, $C_2$ are constants and $\lambda$ is the parameter. By substituting (26) into (25) one obtains the homogeneous system for the constants

$$\left[\lambda^2-\frac{\left(a_3^2-a_2^2\right)}{\left(a_2^2+a_3^2\right)}\left(\omega_0^2+\frac{5g}{a_3}\right)\right]C_1+\lambda\frac{2a_2^2\omega_0^2}{a_2^2+a_3^2}C_2=0$$

$$-\frac{2a_1^2\omega_0^2}{a_1^2+a_3^2}C_1+\left[\lambda^2-\frac{\left(a_3^2-a_1^2\right)}{\left(a_1^2+a_3^2\right)}\left(\omega_0^2+\frac{5g}{a_3}\right)\right]C_2=0 \qquad (27)$$

For the nontrivial solution the determinant of this system must be zero

$$\lambda^4+\frac{2\left[\left(a_3^4+a_1^2a_2^2\right)\omega_0^2-\frac{5g}{a_3}\left(a_3^4-a_1^2a_2^2\right)\right]}{\left(a_3^2+a_1^2\right)\left(a_3^2+a_2^2\right)}\lambda^2+\left(\omega_0^2+\frac{5g}{a_3}\right)^2\frac{a_3^2-a_1^2}{a_3^2+a_1^2}\frac{a_3^2-a_2^2}{a_3^2+a_2^2}=0 \qquad (28)$$

This is the characteristic equation of the problem and has the form $\lambda^4+2p\lambda^2+q=0$. The roots of this equation are

$$\lambda_{1,2}^2=-\left(p\pm\sqrt{p^2-q}\right) \qquad (29)$$

For stability, these roots should be real and negative since then all four roots $\lambda$ are pure imaginaries and therefore the perturbations remain bounded. Now both the roots (29) will be real and negative if $p \pm \sqrt{p^2 - q} > 0$. From this it follows that if the roots are real and negative then one must have

$$p > 0 \quad \text{and} \quad q > 0 \quad \text{and} \quad p > \sqrt{q} \tag{30}$$

Applying (30) to (28) one finds

$$\begin{aligned} &\omega_0^2 > \left(\frac{5g}{a_3}\right)\frac{a_3^4 - a_1^2 a_2^2}{a_3^4 + a_1^2 a_2^2} \qquad \left(a_3^2 - a_2^2\right)\left(a_3^2 - a_1^2\right) > 0 \\ &\omega_0^2 > \left(\frac{5g}{a_3}\right)\frac{a_3^4 - a_1^2 a_2^2 + \sqrt{\left(a_3^4 - a_1^4\right)\left(a_3^4 - a_2^4\right)}}{a_3^4 - a_1^2 a_2^2 - \sqrt{\left(a_3^4 - a_1^4\right)\left(a_3^4 - a_2^4\right)}} \end{aligned} \tag{31}$$

It can be shown that the first condition - except for the case $a_1 = a_2 = a_3$ - gives lower values of $\omega_0$ than the third condition. In this way, one controls stability. By multiplying its numerator and denominator with $a_3^4 - a_1^2 a_2^2 - \sqrt{\left(a_3^4 - a_1^4\right)\left(a_3^4 - a_2^4\right)}$ and taking the square root one obtains the condition for steady stable vertical rotation in the form

$$\boxed{\left|\omega_0\right| > \sqrt{\frac{5g}{a_3}}\,\frac{\sqrt{a_3^4 - a_1^4} + \sqrt{a_3^4 - a_2^4}}{a_1^2 + a_2^2}} \tag{32}$$

This matches Puiseux's solution. Further, the second condition of (31) implies that in addition one must have

$$a_3 > a_1 \quad \text{and} \quad a_3 > a_2 \qquad \text{or} \qquad a_1 > a_3 \quad \text{and} \quad a_2 > a_3 \tag{33}$$

In other words: for stability the vertical semi axis of an ellipsoid should be the largest or the smallest of the ellipsoid's semi axes. Condition (32) is for the case when $a_3$ is the largest, if $a_3$ is the smallest then the vertical rotation is stable for any angular velocity $\omega_0$.

## Appendix.

Bryant or Cardan angles ([5] pp 21-22,[1] pp 608-610 ) are obtained by the consecutive rotations through the angle $\theta_1$ about the axis $\xi_1$, then through the angle $\theta_2$ about intermediary $\xi_2$ and finally through the angle $\theta_3$ about intermediary $\xi_3$ axis. This sequence of rotations leads to the following relation between the body-fixed base and reference base

$$\begin{bmatrix} \hat{\boldsymbol{\varepsilon}}_1 \\ \hat{\boldsymbol{\varepsilon}}_2 \\ \hat{\boldsymbol{\varepsilon}}_3 \end{bmatrix} = \begin{bmatrix} c_2c_3 & c_1s_3 + s_1s_2c_3 & s_1s_3 - c_1s_2c_3 \\ -c_2s_3 & c_1c_3 - s_1s_2s_3 & s_1c_3 + c_1s_2s_3 \\ s_2 & -c_1s_2 & c_1c_2 \end{bmatrix} \begin{bmatrix} \hat{\boldsymbol{e}}_1 \\ \hat{\boldsymbol{e}}_2 \\ \hat{\boldsymbol{e}}_3 \end{bmatrix} \tag{A34}$$

where $c_i = \cos\theta_i$ and $s_i \equiv \sin\theta_i$ $(i = 1,2,3)$. From the relation (A34) it follows that the components of angular velocity in the body fixed base are given by

$$\begin{bmatrix} \omega_1 \\ \omega_2 \\ \omega_3 \end{bmatrix} = \begin{bmatrix} \cos\theta_2 \cos\theta_3 & \sin\theta_3 & 0 \\ -\cos\theta_2 \sin\theta_3 & \cos\theta_3 & 0 \\ \sin\theta_2 & 0 & 1 \end{bmatrix} \begin{bmatrix} \dot{\theta}_1 \\ \dot{\theta}_2 \\ \dot{\theta}_3 \end{bmatrix} \tag{A35}$$

The inversion of the matrix fails for $\theta_2 = \pi/2 + n\pi \ \ (n = 0, \pm1, \pm2, ...)$.